\documentclass[titlepage,12pt]{article}
\usepackage{amssymb,psfig,epsfig,pslatex,psfrag}

\usepackage[latin1]{inputenc}
\usepackage[T1]{fontenc}
\newcommand{\bfsig}{{\mbox{\boldmath $\sigma$}}}

\newcommand{\one}{1\!\mbox{l}}

\newcommand{\bea}{\begin{eqnarray}}
\newcommand{\eea}{\end{eqnarray}}
\begin{document}
\begin{titlepage}
\noindent
DESY 03-004 \hfill January 2003 \\ 
\vspace{0.4cm}
\renewcommand{\thefootnote}{\fnsymbol{footnote}}
\begin{center}
{\LARGE {\bf SUSY QCD corrections to the polarization \\ and spin correlations 
of top quarks \\ produced in $e^+e^-$ collisions}} \\
\vspace{2cm}
{\bf 
A. Brandenburg $^{a,}$\footnote{supported by a Heisenberg fellowship of D.F.G.}and M. Maniatis $^{b}$}
\par\vspace{1cm}
$^a$  DESY-Theorie, D-22603 Hamburg, Germany\\
$^b$  II. Insitut f\"ur Theoretische Physik, Universit\"at Hamburg, Luruper
Chaussee 149, D-22761 Hamburg, Germany
\par\vspace{3cm}
{\bf Abstract:}\\
\parbox[t]{\textwidth}
{We compute the supersymmetric QCD corrections to the polarization 
and the spin correlations of 
top quarks produced above threshold in $e^+e^-$ collisions, 
taking into account
arbitrary longitudinal polarization of the initial beams.}
\end{center}
\vspace{2cm}
%PACS number(s): 12.38.Bx, 13.88.+e, 14.65.Ha\\
\end{titlepage}
\renewcommand{\thefootnote}{\arabic{footnote}}
\setcounter{footnote}{0}
\section{Introduction}
A future linear $e^+e^-$ collider will be an excellent tool 
to search for and investigate extensions of the  Standard
Model (SM) of particle physics \cite{tesla}. 
One particularly attractive
extension of the SM is Supersymmetry (SUSY) \cite{susy}, 
which solves several conceptual problems of the SM.
Apart from  their direct production, also virtual effects of SUSY particles
may lead to observable deviations from the SM expectations. In particular, 
top quark pair production at a linear collider may be a sensitive probe 
of such effects. Very high energy scales are involved in the production and 
decay of top quarks. Moreover, since they decay very quickly, 
the spin of top quarks is not affected by hadronization effects and becomes
an additional observable to probe top quark interactions. At a 
future linear $e^+e^-$ collider,  
the electron (and possibly also the positron) beam may have a substantial
longitudinal polarization, which will be an asset to study top quark spin 
phenomena. We therefore study in this paper the impact of 
virtual effects of SUSY
particles on spin properties of $t\bar{t}$ pairs in $e^+e^-$ collisions.
We restrict ourselves here to the SUSY QCD sector of the Minimal Supersymmetric
Standard Model (MSSM). SUSY QCD corrections to the (spin-summed) differential
cross section for $e^+e^-\to t\bar{t}$ have already been studied quite some 
time ago \cite{DjDrKo93}, and we extend these results by keeping the full
information on the $t\bar{t}$ spin state.
The full MSSM corrections to the spin-summed differential 
cross section have been calculated in \cite{HoSc98}. 

In section 2 we define the spin observables that we calculate in this paper 
and also discuss how they can be measured. Section 3 gives analytic results
for these observables, and section 4 contains numerical results for specific 
choices of the SUSY QCD parameters. In section 5 we present our conclusions.  
\section{Spin observables}\label{sec2}
We consider the reaction
  \begin{equation}
    \label{process} e^+(p_+,\lambda_+)+e^-(p_-,\lambda_-)\to (\gamma^\ast,Z^\ast)\to t(k_t)+
    \bar t(k_{\bar t}) + X,
  \end{equation}
where $\lambda_-$ ($\lambda_+$) denotes the longitudinal 
polarization  of the electron (positron) beam\footnote{
For a right-handed electron (positron), $\lambda_{\mp}=+1$.}.
Within the Standard Model, spin effects of top quarks 
in reaction (\ref{process}) have been analysed
first in ref. \cite{KuReZe86}.  
QCD corrections to the production of top quark pairs, including the 
full information about their spins, can be found in 
refs.~\cite{BrFlUw99,schmidt}. 
Fully analytic results 
for the top quark polarization \cite{epeman1} and a specific spin correlation 
\cite{epeman2} to order $\alpha_s$ are also available. 
\par
The top quark polarization is defined as two times the expectation value
of the top quark spin operator  ${\bf S}_t$.
The operator  ${\bf S}_t$ acts on the tensor product of
the $t$ and $\bar{t}$  spin spaces and is given by 
${\bf S}_t= \frac{\bfsig}{2}\otimes \one $, where
the first (second) factor in the tensor product 
refers to the $t$ ($\bar{t}$) spin space. (The spin operator of the
top antiquark is defined by ${\bf S}_{\bar t}= \one \otimes \frac{\bfsig}{2}$.)
The expectation value is taken with respect to the spin degrees
of freedom of the $t\bar{t}$ sample described 
by a spin density matrix $R$,
i.e.
\bea \label{pol}
{\bf P}_t = 2\,\langle {\bf S}_t\rangle = 
2\frac{{\rm Tr}\, \left[R\, {\bf S}_t\right]}{{\rm Tr}\, R}.
\eea  
For details on the definition and computation of $R$, see e.g. 
\cite{BrFlUw99}.
The polarization of the top antiquark ${\bf P}_{\bar t}$ 
is defined by replacing ${\bf S}_t$ by ${\bf S}_{\bar t}$ in (\ref{pol}).
For top quark pairs produced by CP invariant interactions, we have 
${\bf P}_{\bar t}={\bf P}_{t}$.
The spin correlations between
$t$ and $\bar{t}$ can be calculated by using the matrix
\bea \label{corr} 
C_{ij} = 4\,\langle S_{t,i} S_{\bar{t},j} \rangle = 
4\frac{{\rm Tr}\,\left[ R\,S_{t,i}S_{\bar{t},j}\right]}{{\rm Tr}\, R}.
\eea
Using arbitrary spin quantization axes 
$\hat{\bf{a}}$ and $\hat{\bf{b}}$  for the 
$t$ and $\bar{t}$ spins, the spin correlation with respect to these axes 
is given by
\bea\label{corr}
c(\hat{\bf{a}},\hat{\bf{b}})=\frac{\hat{a}_iC_{ij}\hat{b}_j
-({\bf P}_{t}\cdot\hat{\bf{a}})({\bf P}_{\bar{t}}\cdot\hat{\bf{b}})}
{\sqrt{1-({\bf P}_{t}\cdot\hat{\bf{a}} \!\!\!\! \phantom{\hat{\bf{b}}})^2}
 \sqrt{1-({\bf P}_{\bar{t}}\cdot\hat{\bf{b}})^2}}.
\eea
The directions
$\hat{\bf a}$, $\hat{\bf b}$ can be 
chosen arbitrarily. Different choices will yield 
different values for the spin correlation $c(\hat{\bf{a}},\hat{\bf{b}})$.
The spin properties of the top quarks and antiquarks can be measured
by analysing the angular distributions of the $t$ and $\bar{t}$ decay
products. For example, if both $t$ and $\bar{t}$ decay semileptonically,
$t\to b\ell^+\nu_\ell,\ \bar{t}\to \bar{b}\ell\,'^-\bar{\nu}_{\ell'}$, 
the following
double differential lepton angular distribution is sensitive to
the $t\bar{t}$ spin state:
\begin{eqnarray}
\frac{1}{\sigma}\frac{d^2\sigma}{d\cos\theta_+ d\cos\theta_-}=
\frac{1}{4} (1 +
{\rm B}_1\cos\theta_+
+ {\rm B}_2\cos\theta_- 
-{\rm C}\cos\theta_+ \cos\theta_-)\,\, ,
\label{eq:ddist1}
\end{eqnarray}
with $\sigma$ being the cross section for the channel under consideration.
In Eq.~(\ref{eq:ddist1})  $\theta_+$ ($\theta_-$) denotes the angle between the
direction of flight of the lepton $\ell^+$ ($\ell\,'^-$) in the $t$ ($\bar{t}$)
rest frame and the chosen spin quantization axis
$\hat{\bf a}$ ($\hat{\bf b}$). The coefficients $B_{1,2}$ and $C$ are related
to the mean (averaged over the scattering angle) $t$ ($\bar{t}$) polarization 
and spin correlation projected onto the directions $\hat{\bf a}$ 
and $\hat{\bf b}$. Using the double pole approximation \cite{Stu91}
for the $t$ and $\bar{t}$ propagators, 
one obtains for the so-called factorizable contributions 
\cite{BeBeCh99,BeBrSiUw01}
\bea \label{bbc}
B_1 &=& \kappa_+\overline{{\bf P}_t\cdot \hat{\bf a}},\nonumber \\
B_2 &=& -\kappa_-\overline{{\bf P}_{\bar{t}}\cdot \hat{\bf b}}, \nonumber \\
C  &=& \kappa_+\kappa_-\overline{\hat{a}_iC_{ij}\hat{b}_j},
\eea
where the overline indicates the average over the scattering angle, e.g.
\bea\label{proj}
\overline{{\bf P}_t\cdot \hat{\bf a}}= 
2\frac{\int_{-1}^1dy{\rm Tr}\, \left[R\, {\bf S}_t\cdot \hat{\bf a}
\right]}{\int_{-1}^{1}dy{\rm Tr}\, R},
\eea
etc., where  $y$ is the cosine of the top quark scattering 
angle. In (\ref{bbc}), $\kappa_\pm$ is the spin analysing
power of the charged lepton $\ell^\pm$. At leading order,
$\kappa_\pm = +1$. QCD corrections to this result are at the per mill
level \cite{CzJeKu91}. SUSY QCD corrections to the spin 
analysing power $\kappa_\pm$ are exactly zero \cite{BrMa02}.
\section{Analytic results}\label{sec3}
We now turn towards the calculation of the SUSY QCD corrections
to the polarization and spin correlations of top quark pairs produced
in $e^+e^-$ collisions. These corrections directly determine
the SUSY QCD corrections to the double lepton distribution (\ref{eq:ddist1})
within the double pole approximation, since the corrections
to  the LO result $\kappa_\pm=+1$ are exactly zero and the 
non-factorizable contributions
due to SUSY particles also vanish within that approximation.

The amplitude for reaction (\ref{process}) including SUSY QCD corrections
may be written as follows:
\bea\label{amplitude}
iT_{fi} &=& i\frac{4\pi\alpha}{s}\Bigg\{\chi(s)\bar{v}(p_+)
\left(g_V^e\gamma_{\mu}-g_A^e\gamma_{\mu}\gamma_5\right)u(p_-)H_Z^{\mu}
-\bar{v}(p_+)\gamma_{\mu}u(p_-)H_{\gamma}^{\mu}
\Bigg\},
\eea
where $g_V^e = -\frac{1}{2} + 2 \sin^2\vartheta_W$, and
$g_A^e =-\frac{1}{2}$, with $\vartheta_W$ denoting the 
weak mixing angle.
The function
$\chi$ is given by
  \begin{equation}
    \label{chi}
    \chi(s) = \frac{1}{4\sin^2\vartheta_W\cos^2\vartheta_W}\,
    \frac{s}{s-m_Z^2},
  \end{equation}
where $m_Z$ stands for the mass of the Z boson. We neglect the Z width,
since we work at lowest order in the electroweak coupling and the
c.m. energy is far above $m_Z$. 
The hadronic currents have a formfactor decomposition as follows:
\bea\label{hc}
H_{Z,\gamma}^{\mu}= \bar{u}(k_t)\left[
V_{Z,\gamma}\gamma^{\mu}-A_{Z,\gamma}\gamma^{\mu}\gamma_5
+S_{Z,\gamma}\frac{\left(k_t-k_{\bar{t}}\right)^{\mu}}{2m_t}
\right]v(k_{\bar{t}})
\eea
with
\bea\label{ff}
V_{Z,\gamma}&=&V^0_{Z,\gamma} + V^1_{Z,\gamma},\nonumber \\
A_{Z,\gamma}&=&A^0_{Z,\gamma} + A^1_{Z,\gamma}.
\eea
In (\ref{ff}),
$V^0_{\gamma}=Q_t$, where $Q_t$ denotes 
the electric charge of the top quark in units of $e=\sqrt{4\pi\alpha}$, 
$A^0_{\gamma}=0$, and
$V^0_{Z}= g_V^t=\frac{1}{2} - \frac{4}{3} \sin^2\vartheta_W$, 
$A^0_{Z}=g_A^t=\frac{1}{2}$ are 
the tree level vector- and the axial-vector couplings of the
top quark to the Z boson.

The one-loop SUSY QCD contributions 
to the different form factors are denoted by 
$V^1_{\gamma,Z}$, $A^1_{\gamma,Z}$ and $S_{\gamma,Z}$. 
Scalar and pseudoscalar couplings proportional to
$\left(k_t+ k_{\bar{t}}\right)^{\mu}$ and 
$\left(k_t+ k_{\bar{t}}\right)^{\mu}\gamma_5$ have been neglected in 
(\ref{amplitude}), since they induce contributions proportional to
the electron mass.
In addition CP violating formfactors proportional to 
$\left(k_t- k_{\bar{t}}\right)^{\mu}\gamma_5$ are possible in SUSY QCD 
through a complex phase in the squark mass matrices. 
In \cite{Sc02} it has been shown that the dependence
of the cross section on these phases is weak and that CP odd asymmetries
are typically of the order of $10^{-3}$. 
We therefore set these phases
to zero in the following.
To make this paper 
self-contained we list the form factors $V^1_{\gamma,Z}$, 
$A^1_{\gamma,Z}$ and $S_{\gamma,Z}$ in the appendix. 
We have performed an analytic comparison to the
corresponding results in \cite{DjDrKo93} and found complete agreement.
 
We define
\begin{eqnarray}\label{wcouplings}
f_{LL(LR)}&=&-Q_t+\chi(g_V^e+g_A^e)(g_V^t\pm g_A^t),\nonumber \\
f_{RR(RL)}&=&-Q_t+\chi(g_V^e-g_A^e)(g_V^t\mp g_A^t),
\end{eqnarray}
and
\begin{eqnarray}
P_\pm = 1-\lambda_-\lambda_+\pm(\lambda_- - \lambda_+).
\end{eqnarray}

The electroweak couplings that enter
the Born results are then given by 
\begin{eqnarray}
g_{VV}^\pm&=&\frac{1}{8}\left[P_+(f_{RR}+f_{RL})^2\pm 
P_-(f_{LL}+f_{LR})^2\right],\nonumber \\
g_{AA}^\pm&=&\frac{1}{8}\left[P_+(f_{RR}-f_{RL})^2\pm 
P_-(f_{LL}-f_{LR})^2\right],\nonumber \\
g_{VA}^\pm&=&\frac{1}{8}\left[P_+(f_{RR}^2-f_{RL}^2)\pm 
P_-(f_{LL}^2-f_{LR}^2)\right].
\end{eqnarray}

Likewise, defining 

\begin{eqnarray}
g_{LL(LR)}&=& \chi  (g_V^e+g_A^e)(V^1_Z\pm A^1_Z)-(V^1_{\gamma}
\pm A^1_{\gamma} )
\nonumber \\
g_{RR(RL)}&=& \chi (g_V^e-g_V^e)(V^1_Z\mp A^1_Z)
-(V^1_{\gamma}\mp A^1_{\gamma} ) 
,\nonumber \\               
s_{L(R)}&=& \chi (g_V^e\pm g_A^e)S_Z-S_{\gamma},      .
\end{eqnarray}

the SUSY QCD contributions may be written in terms of the 
following quantities:

\begin{eqnarray}
h_{VV}^\pm&=&\frac{1}{8}\left[P_+(f_{RR}+f_{RL})(g_{RR}+g_{RL})\pm 
P_-(f_{LL}+f_{LR})(g_{LL}+g_{LR})\right],\nonumber \\
h_{AA}^\pm&=&\frac{1}{8}\left[P_+(f_{RR}-f_{RL})(g_{RR}-g_{RL})\pm 
P_-(f_{LL}-f_{LR})(g_{LL}-g_{LR})\right],\nonumber \\
{\rm Re\ }h_{VA}^\pm&=&\frac{1}{8}{\rm Re\ }
\left[P_+(f_{RR}g_{RR}-f_{RL}g_{RL})\pm 
P_-(f_{LL}g_{LL}-f_{LR}g_{LR})\right],\nonumber \\
{\rm Im\ }h_{VA}^\pm&=&\frac{1}{8}{\rm Im\ }\left[P_+(f_{RL}g_{RR}-f_{RR}g_{RL})\pm 
P_-(f_{LR}g_{LL}-f_{LL}g_{LR})\right],\nonumber \\
s_{V}^{\pm}&=&\frac{1}{4}\left[P_+(f_{RR}+f_{RL})s_R\pm 
P_-(f_{LL}+f_{LR})s_L\right],\nonumber \\
s_{A}^{\pm}&=&-\frac{1}{4}\left[P_+(f_{RR}-f_{RL})s_R\pm 
P_-(f_{LL}-f_{LR})s_L\right].
\end{eqnarray}

It is convenient to write the results 
in terms of the electron and top quark
directions
$\hat{\bf p}$ and $\hat{\bf k}$ defined in the c.m. system, the
cosine of the scattering angle $y=\hat{\bf p}\cdot\hat{\bf k}$, the 
scaled top quark mass $r=2m_t/\sqrt{s}$, and the top quark velocity
$\beta=\sqrt{1-r^2}$. 

\par
The differential cross section including the SUSY QCD corrections reads:
\bea\label{sigma}
\frac{d\sigma}{dy}&=& 
\frac{d\sigma^{0}}{dy}+\frac{d\sigma^1}{dy}
=\sigma_{\rm pt}\frac{3N_C\beta}{8}
\Bigg\{\left[2-\beta^2 (1-y^2)\right]\left(g^{+}_{VV}+
2{\rm Re\ }h^{+}_{VV}\right)
\nonumber \\
&+&\beta^2 (1+y^2 )\left(g^{+}_{AA}+2{\rm Re\ }h^{+}_{AA}\right)
+4\beta y \left(g^{+}_{VA}+2{\rm Re\ }h^{+}_{VA}\right)-2\beta^2(1-y^2) 
{\rm Re\ }s^{+}_V
\Bigg\},
\eea
where 
\bea
\sigma_{\rm pt} = \frac{4\pi\alpha^2}{3s},
\eea
and $d\sigma^{0}/dy$ is obtained by setting 
$h^{+}_{VV}=h^{+}_{AA}=h^{+}_{VA}=s^{+}_V=0$.
We further introduce a vector perpendicular to ${\bf k}$ in the
production plane
${\bf k}^{\perp}=\hat{\bf p}-y\hat{\bf k}$ and a vector normal to this plane,
${\bf n}=\hat{\bf p}\times \hat{\bf k}$. 
The top quark polarization including the SUSY QCD corrections 
is equal to the top antiquark
polarization and  reads:
\bea\label{polres}
{\bf P}_t&=&{\bf P}_t^0+{\bf P}^1_t\nonumber \\
&=&\sigma_{\rm pt}\frac{3N_C\beta}{4}
\Bigg\{ \left[\beta (1+y^2 )\left(g^{-}_{VA}+2{\rm Re\ }h^{-}_{VA}\right)
+y\left(g^{-}_{VV}+2{\rm Re\ }h^{-}_{VV}\right)
+\beta^2 y  \left(g^{-}_{AA}+2{\rm Re\ }h^{-}_{AA}\right)\right]
 \hat{\bf k}\nonumber\\
 &+&
r\left[\beta y \left(g^{-}_{VA}+2{\rm Re\ }h^{-}_{VA}\right)
+g^{-}_{VV}+2{\rm Re\ }h^{-}_{VV}
-\frac{\beta^2}{r^2}\left({\rm Re\ }s^-_V-\beta y {\rm Re\ }s^{-}_A\right)
\right]{\bf k}^{\perp}\nonumber \\
&+&
\left[2\beta r {\rm Im\ }h^{+}_{VA}
+\frac{\beta^2}{r}
\left(y {\rm Im\ }s_V^+ -\beta{\rm Im\ }s_A^+\right)\right]{\bf n}
\Bigg\}
\left(\frac{d\sigma^0}{dy}\right)^{-1}-{\bf P}_t^0
\frac{d\sigma^1}{dy}
\left(\frac{d\sigma^0}{dy}\right)^{-1}.
\eea
For the matrix $C_{ij}$ defined in (\ref{corr}) we find
\bea\label{corrres}
C_{ij}&=&C_{ij}^0+ C_{ij}^1
=  \frac{1}{3}\delta_{ij}\left[1+\frac{d\sigma^1}{dy}
\left(\frac{d\sigma^0}{dy}\right)^{-1}\right] \nonumber \\
&+& \sigma_{\rm pt}\frac{3N_C\beta}{4}
\left(\frac{d\sigma^0}{dy}\right)^{-1}
\Bigg\{\left[g^{+}_{VV}+2{\rm Re\ }h^{+}_{VV}-
\beta^2 \left(g^{+}_{AA}+2{\rm Re\ }h^+_{AA}\right)\right] 
\left[k^{\perp}_i k^{\perp}_j-\frac{1}{3}\delta_{ij}
(1-y^2)\right]
\nonumber \\ &+&
 \Big[\left(y^2 +\beta^2(1-y^2)\right)\left(g^{+}_{VV}
+2{\rm Re\ }h^{+}_{VV}\right)+
\beta^2 y^2 \left(g^{+}_{AA}+2{\rm Re\ }h^{+}_{AA}\right)
+2\beta y \left(g^{+}_{VA}+2{\rm Re\ }h^{+}_{VA}\right)
\nonumber \\
&+&2\beta^2(1-y^2){\rm Re\ }s_V^+\Big]
\left[\hat{k}_i\hat{k}_j-\frac{1}{3}\delta_{ij}\right]
\nonumber \\  
&+&
r \left[y \left(g^{+}_{VV}+2{\rm Re\ }h^{+}_{VV}\right)
+ \beta \left(g^{+}_{VA}+2{\rm Re\ }h^{+}_{VA}\right)
-\frac{\beta^2}{r^2}\left( y{\rm Re\ }s_V^+ - \beta {\rm Re\ }
s_A^+\right)\right]
\left[k^{\perp}_i\hat{k}_j+k^{\perp}_j\hat{k}_i\right]
\nonumber \\ &+&
2\beta{\rm Im\ }h_{VA}^{-}
\left[k^{\perp}_i n_j+k^{\perp}_j n_i\right]
+\beta\left[2 y r {\rm Im\ }h_{VA}^{-}
+\frac{\beta}{r}\left({\rm Im\ }s_V^{-}-\beta y {\rm Im\ }s_A^{-}
\right)\right]\left[\hat{k}_i n_j+\hat{k}_j n_i\right]
\nonumber \\
&-& {C}_{ij}^0
\frac{d\sigma^1}{dy}
\left(\frac{d\sigma^0}{dy}\right)^{-1}
\Bigg\}.
\eea
The Born results ${\bf P}_t^0$ and $C_{ij}^0$ are 
obtained from (\ref{polres}) and  (\ref{corrres}) by setting
$h^{\pm}_{VA}=h^{-}_{VV}=h^{-}_{AA}=s^{\pm}_{V}=s^{\pm}_{A}
=d\sigma^1/{dy}=0$.

For fully polarized electrons (or positrons) a so-called `optimal
spin basis' can be constructed. This is an axis $\hat{\bf d}$
with respect to 
which the $t$ and $\bar{t}$
spins are  100\% correlated  at the tree level in the Standard Model 
for any velocity and scattering
angle \cite{PaSh96}. This axis $\hat{\bf d}$ is the solution of the equation
\bea
\hat{d}_iC_{ij}^0\hat{d}_j=1.
\eea
One gets
\bea\label{d}
\hat{\bf d}= x\hat{{\bf k}}+\sqrt{1-x^2}\hat{{\bf k}}^{\perp},
\eea
with $x\in [-1,1]$ only if either $P_+=0$ or $P_-=0$. 
For $P_+=0$, which
can be realized with left-handed electrons ($\lambda_-=-1$), one finds
\bea\label{x}
x=-\frac{f_{LL}(\beta+y)+f_{LR}(y-\beta)}{\left[(1+y\beta)^2f_{LL}^2
+(1-y\beta)^2f_{LR}^2+2(y^2\beta^2+1-2\beta^2)f_{LL}f_{LR}
\right]^{1/2}}.
\eea 
For right-handed electrons, the optimal basis is obtained
by replacing $f_{LL}\to f_{RR}, \ f_{LR}\to f_{RL}$ in Eq. (\ref{x}).
Note that at threshold 
$\hat{\bf d}{\buildrel
\beta\to 0\over \longrightarrow} \hat{\bf p},$
i.e. the optimal basis at threshold is defined by the direction of the
beam, while in the high-energy limit $\hat{\bf d}{\buildrel
\beta\to 1\over \longrightarrow} \hat{\bf k}$, i.e. the optimal basis
coincides with the helicity basis.
By analytically evaluating $\hat{d}_iC^1_{ij}\hat{d}_j$
we find that the virtual SUSY QCD corrections to the $t\bar{t}$ 
spin correlations in the optimal basis are exactly zero. 
\section{Numerical results}
In this section we present numerical results for the SUSY QCD
corrections to the top quark polarization and $t\bar{t}$ spin correlations.
We also include a discussion of  the corrections to the differential
cross section and compare our results to the literature. 

We take into account the effects 
of mixing of the chiral components of the top squark.
The stop mass matrix can be expressed in terms of MSSM
parameters as follows:
\begin{eqnarray}
{\cal M}_{\tilde{t}}^2 \!\!\! &=& \!\!\! \left(\begin{array}{cc} M_{\tilde{Q}}^2+m_t^2+m_Z^2(\frac{1}{2}-Q_ts_W^2)\cos 2\beta& m_t(A_t-\mu\cot\beta)\\ m_t(A_t-\mu\cot\beta)&
\!\!\!\!\!\!\!\!\!\!\!\!\!\!\!\!  M_{\tilde{U}}^2+m_t^2+m_Z^2Q_ts_W^2\cos 2\beta\end{array}\right),
% \nonumber \\ && \nonumber \\ && \nonumber \\
%{\cal M}_{\tilde{b}}^2 \!\!\! &=& \!\!\! \left(\begin{array}{cc} M_{\tilde{Q}}^2+m_b^2-m_Z^2(\frac{1}{2}+Q_bs_W^2)\cos 2\beta&m_b(A_b-\mu\tan\beta)\\ m_b(A_b-\mu\tan\beta)&
%\!\!\!\!\!\!\!\!\!\!\!\!\!\!\!\! M_{\tilde{D}}^2+m_b^2+m_Z^2Q_bs_W^2\cos 2\beta
%\end{array}\right)
\end{eqnarray}
where $M_{\tilde{Q}},\ M_{\tilde{U}}$ are the soft SUSY-breaking
parameters for the squark doublet $\tilde{q}_L$ ($q=t,b$) 
and the top squark singlet
$\tilde{t}_R$, respectively. Further, $A_{t}$
is the stop soft SUSY-breaking trilinear coupling,
and $\mu$ is the SUSY-preserving bilinear Higgs coupling.
The ratio of the two Higgs vacuum expectation values is given by $\tan\beta$,
and we use the abbreviation $s_W=\sin\theta_W$. 
The squared physical masses of the stops
are the eigenvalues of the above matrix.
In order to simplify the discussion, we set
$\tan\beta=1$ for all following results.
Further, we assume that the sbottom mass matrix is diagonal with
degenerate mass eigenvalues, ${\cal M}_{\tilde{b}}^2=
{\rm diag}(m^2_{\tilde{b}},m^2_{\tilde{b}})$.
Neglecting $m_b$ in the sbottom mass matrix this leads to $M_{\tilde{Q}}= 
m_{\tilde{b}}$, and
the stop mass matrix simplifies
under the above assumptions to
\begin{eqnarray}
{\cal M}_{\tilde{t}}^2 &=&
\left(\begin{array}{cc} m_{\tilde{b}}^2+m_t^2& m_tM_{LR}\\ m_tM_{LR}&
M_{\tilde{U}}^2+m_t^2
\end{array}\right),
\end{eqnarray}
with $M_{LR}= A_t-\mu$.
The stop mass eigenstates are obtained from  the chiral states by
a rotation:
\begin{eqnarray}\label{mix} 
\left(\begin{array}{l}\tilde{t}_1 \\  \tilde{t}_2 \end{array}\right)=
\left(\begin{array}{cc} \cos\theta_{\tilde{t}} & 
\sin\theta_{\tilde{t}} \\ -\sin\theta_{\tilde{t}} &
\cos\theta_{\tilde{t}} 
\end{array}\right)\left(\begin{array}{l}\tilde{t}_L \\  
\tilde{t}_R \end{array}\right).
\end{eqnarray}
Maximal mixing ($\theta_{\tilde{t}}=\frac{\pi}{4}$ and
$M_{LR}\not =0$)
corresponds to $M_{\tilde{U}}^2=m_{\tilde{b}}^2$.
The latter relation will also be assumed for $M_{LR}=0$, leading to
the following stop mass eigenvalues
\footnote{Note that by fixing $\theta_{\tilde{t}}=\frac{\pi}{4}$ the 
light stop can be either $\tilde{t}_{1}$
or $\tilde{t}_{2}$ depending on the sign of $M_{LR}$.}:
\begin{eqnarray} \label{mixsimp}
m_{\tilde{t}_{1,2}}=\sqrt{m_{\tilde{b}}^2+m_t^2\pm m_t M_{LR}}.
\end{eqnarray}
Note that we use here the same set of assumptions on the squark 
mass matrices
as we did in our study of the SUSY QCD corrections in the decay
of polarized top quarks \cite{BrMa02}. 
Further we use  $\sin^2{\theta_W}= 0.2236$, $\alpha_s =$~0.11, 
and we set the top mass to $m_t =$~174~GeV and
the sbottom mass that enters  Eq.~(\ref{mixsimp}) 
to $m_{\tilde{b}}=$~100~GeV.
%
%total cross section

Fig.~1 shows the relative SUSY QCD correction $\sigma^1/\sigma^0$
to the total cross section for $e^+e^-\to t\bar{t}$ with unpolarized 
beams at $\sqrt{s}=500$ GeV
as a function of  the mixing parameter $M_{LR}$,
where $\sigma^0$ and $\sigma^1$ are obtained from Eq.~(\ref{sigma}) by
integrating over $y$. Shown are the
relative corrections for two different gluino masses, namely
$m_{\tilde{g}}=$ 150 GeV and $m_{\tilde{g}}=$ 250 GeV.
%
% total cross section depending on MLR
%
\begin{figure} \label{totalfig}
\begin{center}
\psfrag{MLR}{\small $M_{LR}$ [GeV]}
\psfrag{delta}{\small $\frac{\sigma^1}{\sigma^0}$ [\%]}
\psfrag{mgl250}{\small $m_{\tilde{g}}=$ 250 GeV}
\psfrag{mgl150}{\small $m_{\tilde{g}}=$ 150 GeV}
\psfrag{textM2}{\small $(\sqrt{s}=$ 500 GeV, $m_{\tilde{b}}=$ 100 GeV $)$}       
\includegraphics[angle=-90,width=8cm,clip]{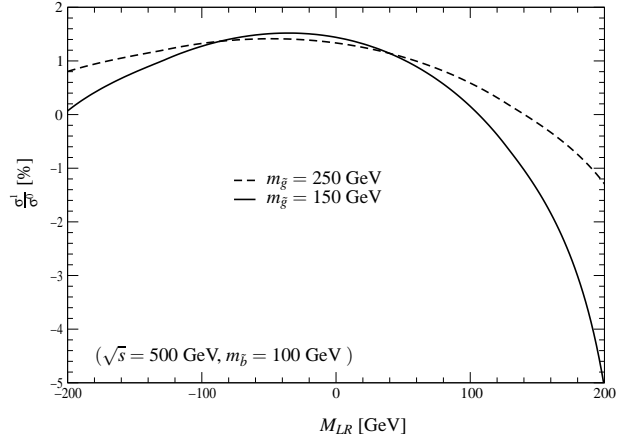}
\caption{Relative correction to the total cross section as a function of the
mixing parameter $M_{LR}$ for a fixed mixing angle
$\theta_{\tilde{q}}=\pi/4$. 
Shown is the correction for two different gluino masses
$m_{\tilde{g}}=$~150 GeV and $m_{\tilde{g}}=$~250 GeV.}
\label{totalfig}
\end{center}
\end{figure}
%%%%%%%%%%%%%%%
For a large mixing parameter $M_{LR}$ and a small gluino mass of  
$m_{\tilde{g}}=$~150 GeV we find a large negative correction. 
The correction
decreases as the gluino mass increases. A mixing parameter of 
$M_{LR} =$~200~GeV corresponds to a light stop mass of 
$m_{\tilde{t}_2}=$~74 GeV, which is above the current experimental
lower limit \cite{aleph}. With our choice of the masses, we are 
far away from the threshold singularity at $m_t=m_{\tilde{g}}+m_{\tilde{t}}$,
where a more sophisticated calculation is necessary.

Fig.~2 shows the
differential cross section $d\sigma/dy$, again for two different
gluino masses 
$m_{\tilde{g}}=$ 150 GeV and $m_{\tilde{g}}=$~250~GeV, 
and for the cases of 
'no mixing' ($M_{LR}=0$) 
and 'mixing' ($M_{LR}=200$ GeV and $\theta_{\tilde{q}}=\pi/4$), again 
at $\sqrt{s}=$~500~GeV.
\begin{figure} \label{figcosbeta}
\begin{center}
\psfrag{cosbeta}{\small $y$}
\psfrag{delta}{\small $\frac{d\sigma^{1}/dy}
{d\sigma^0/dy}$ [\%]}
\psfrag{mgl250}{\small $m_{\tilde{g}}=$ 250 GeV}
\psfrag{mgl150}{\small $m_{\tilde{g}}=$ 150 GeV}
%\psfrag{textM2}{\small $(\sqrt{s}=$ 500 GeV, $m_{\tilde{b}}=$ 100 GeV $)$}       
\includegraphics[angle=-90,width=8cm,clip]{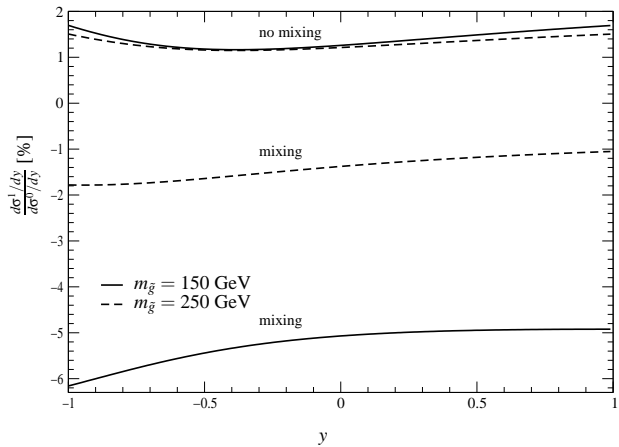}
\caption{Relative correction to the differential cross section $d\sigma/dy$ 
at $\sqrt{s}=500$ GeV for the cases of no mixing 
($M_{LR}=0$)
and mixing
($\theta_{\tilde{q}}=\pi/4$ and $M_{LR}=$~200~GeV).}
\end{center}
\end{figure}
We have compared our results for $\sigma$ and $d\sigma/dy$
with \cite{DjDrKo93} and found agreement with their Fig.~3 (no mixing case), 
while we disagree with the results depicted in Fig.~4 
($\sigma$ and forward-backward asymmetry with stop mixing).
We have also compared our results including the mixing 
with \cite{HoSc98,scha_pr} and find complete agreement.

We now turn towards the discussion of the SUSY QCD corrections to
the $t\bar{t}$ spin properties.

In Fig.~3 we investigate the expectation value of the 
top spin operator as a function of the centre-of-mass energy. 
We have computed the average projected polarization 
defined in Eq.~(\ref{proj}) for three choices of the quantization
axis $\hat{\bf a}$, namely for $\hat{\bf a}=\hat{\bf k}$
(flight direction of the top), 
for $\hat{\bf a}=\hat{\bf p}$ (electron beam direction), 
and for $\hat{\bf a}=\hat{\bf n}$ (normal to the event plane).
These quantities are shown in three different plots, where thin curves
correspond to the tree level results and the thick curves are the relative
corrections in percent. 
The corrections 
are shown for the case of  mixing
($\theta_{\tilde{q}}=\pi/4$ and $M_{LR}=$~200~GeV) and
a gluino mass of $m_{\tilde{g}}=$~150~GeV.
For the polarizations of the initial beams we choose $\lambda_+=0$ 
and consider the three cases $\lambda_-=-1,0,+1$.
The projection of the top quark polarization onto $\hat{\bf n}$  vanishes
at tree level, and thus we only show the contribution 
from SUSY QCD absorptive parts in percent. 
In all cases SUSY QCD effects change the tree level 
results by less than  1\% and vanish at threshold. 
\begin{figure} \label{fig_proj}
\begin{center}
% top
\psfrag{sqrts}{\small $\sqrt{s}$ [GeV]}
\psfrag{proj}{\small $\overline{ {\bf P}_t^0 {\bf \hat{k}}}$ (thin)}
\psfrag{delta}
{\small $\overline{ {\bf P}_t^1 {\bf \hat{k}}}
{\bf /} 
\overline{ {\bf P}_t^0 {\bf \hat{k}}}$ 
[\%] (thick)}
\psfrag{hel+1}{\small $\lambda_-= +$1}
\psfrag{hel0}{\small $\lambda_-=$ 0}
\psfrag{hel-1}{\small $\lambda_-=-$1}                                              
\includegraphics[angle=-90,width=8cm,clip]{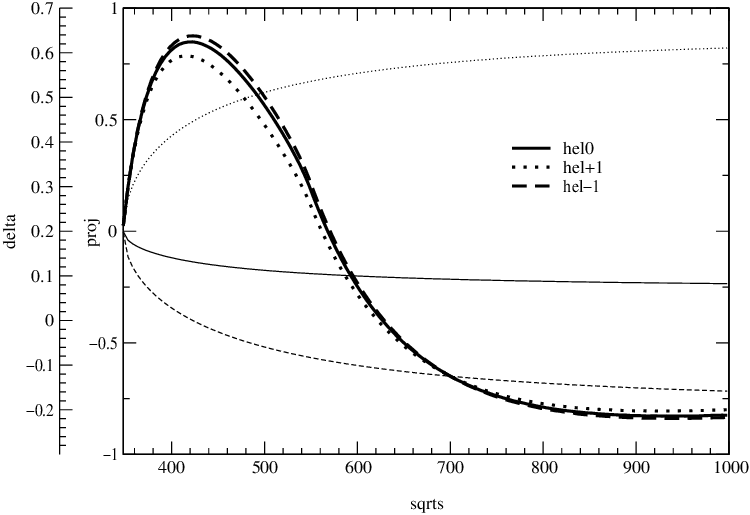}
%
% elc
\psfrag{sqrts}{\small $\sqrt{s}$ [GeV]}
\psfrag{proj}{\small $\overline{ {\bf P}_t^0 {\bf \hat{p}}}$ (thin)}
\psfrag{delta}
{\small $\overline{ {\bf P}_t^1 {\bf \hat{p}}}
{\bf /} 
\overline{ {\bf P}_t^0 {\bf \hat{p}}}$
[\%] (thick)}
\psfrag{hel+1}{\small $\lambda_-= +$1}
\psfrag{hel0}{\small $\lambda_-=$ 0}
\psfrag{hel-1}{\small $\lambda_-=-$1}                                              
\includegraphics[angle=-90,width=8cm,clip]{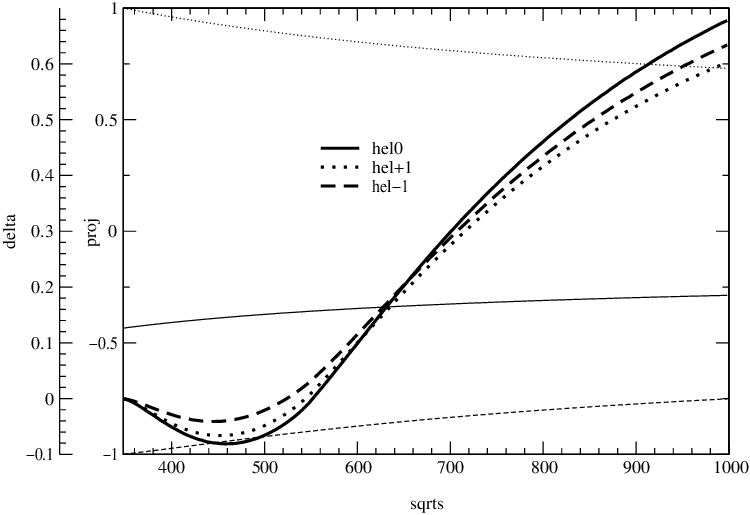}
%
% nor
\psfrag{sqrts}{\small $\sqrt{s}$ [GeV]}
\psfrag{delta}{\small $\overline{ {\bf P}_t {\bf \hat{n}}}$ [\%]}
\psfrag{hel+1}{\small $\lambda_-= +$1}
\psfrag{hel0}{\small $\lambda_-=$ 0}
\psfrag{hel-1}{\small $\lambda_-=-$1}                                              
\includegraphics[angle=-90,width=8cm,clip]{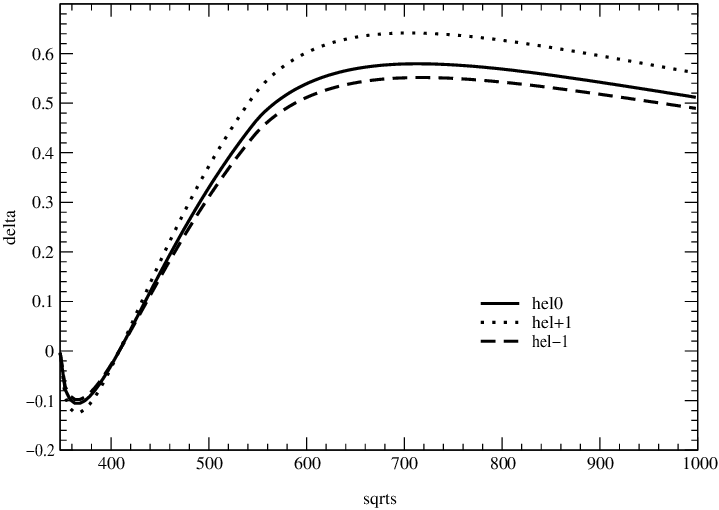}
\caption{Average projected top quark polarization 
$\overline{ {\bf P}_t {\bf \hat{a}}}$ defined in (\ref{proj}) 
for the choices
${\bf \hat{a}}={\bf \hat{k}}$
(top), ${\bf \hat{a}}={\bf \hat{p}}$ (middle), 
${\bf \hat{a}}={\bf \hat{n}}$ (bottom) as a function of 
the centre-of-mass energy. In
each plot we show the tree level results 
(thin lines) and the relative corrections in
percent (thick lines) for unpolarized positrons and the three cases
$\lambda_-=-1,0,+1$.}
\end{center}
\end{figure}

In Fig.~4 we show the averaged spin correlations
$\overline{\hat{a}_iC_{ij}\hat{b}_j}$ for the choices
$\hat{\bf a}=\hat{\bf b}=\hat{\bf k}$ (helicity correlation),
$\hat{\bf a}=\hat{\bf b}=\hat{\bf p}$ (beamline correlation),
and $\hat{\bf a}=\hat{\bf p},\ \hat{\bf b}=\hat{\bf k}$, for the same choice
of parameters as in Fig.~3.
Again the SUSY QCD correction are tiny. Fig.~5 shows the correlations
for the choices $\hat{\bf a}=\hat{\bf k},\ \hat{\bf b}=\hat{\bf n}$ and
$\hat{\bf a}=\hat{\bf p},\ \hat{\bf b}=\hat{\bf n}$. The first of these
two choices of spin quantization axes leads to SUSY QCD effects slightly 
larger than 1\% around c.m. energies of 700 GeV and for a fully polarized
electron beam.

\begin{figure} \label{fig_corr4a4a}
\begin{center}
\psfrag{sqrts}{\small $\sqrt{s}$ [GeV]}
\psfrag{proj}{\small $\overline{\hat{k}_i C_{ij}^0 \hat{k}_j}$ (thin)}
\psfrag{delta}
{\small $ \overline{\hat{k}_i C_{ij}^1 \hat{k}_j}
{\bf /}
\overline{\hat{k}_i C_{ij}^0  \hat{k}_j} $ [\%] (thick)}
\psfrag{hel+1}{\small $\lambda_-= +$1}
\psfrag{hel0}{\small $\lambda_-=$ 0}
\psfrag{hel-1}{\small $\lambda_-=-$1}                                              
\includegraphics[angle=-90,width=8cm,clip]{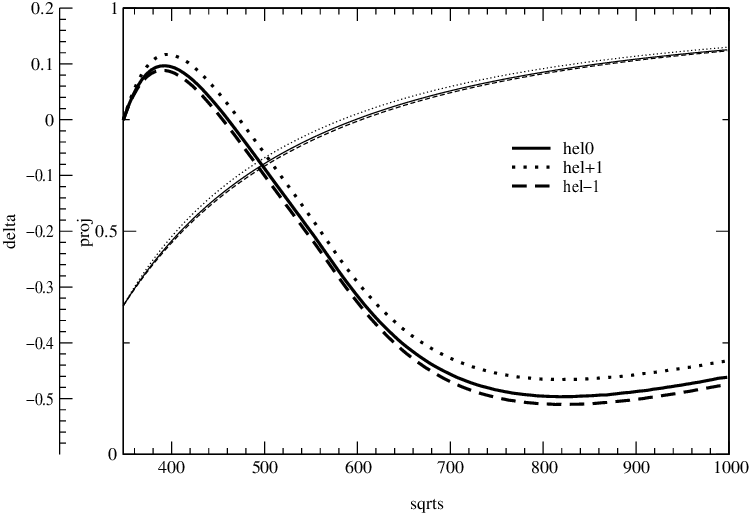}
%
%
% 5a5a
\psfrag{proj}{\small $\overline{\hat{p}_i C_{ij}^0 \hat{p}_j}$ (thin)}
\psfrag{delta}
{\small $ \overline{\hat{p}_i C_{ij}^1 \hat{p}_j}
{\bf /}
\overline{\hat{p}_i C_{ij}^0  \hat{p}_j} $ [\%] (thick)}
\includegraphics[angle=-90,width=8cm,clip]{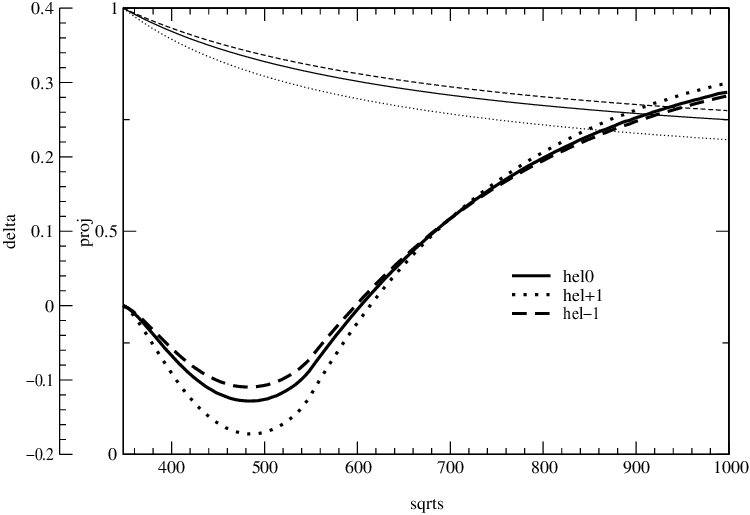}
%
%
% 4a5a
\psfrag{proj}{\small $\overline{\hat{p}_i C_{ij}^0 \hat{k}_j}$ (thin)}
\psfrag{delta}
{\small $ \overline{\hat{p}_i C_{ij}^1 \hat{k}_j}
{\bf /}
\overline{\hat{p}_i C_{ij}^0  \hat{k}_j} $ [\%] (thick)}
\includegraphics[angle=-90,width=8cm,clip]{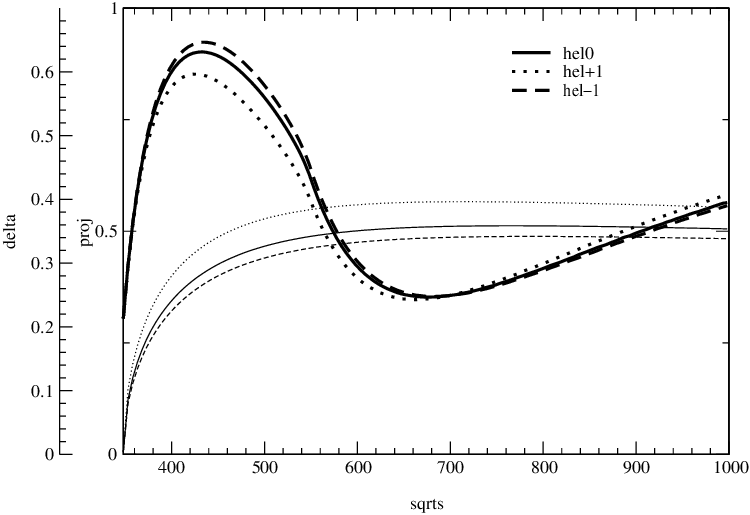}
\caption{Same as Fig.~3, but for the quantities 
$\overline{\hat{k}_i C_{ij}\hat{k}_j}$ (top),
$\overline{\hat{p}_i C_{ij}\hat{p}_j}$ (middle), and
$\overline{\hat{p}_i C_{ij}\hat{k}_j}$ (bottom).}
\end{center}
\end{figure}

\begin{figure}
\begin{center}
\psfrag{sqrts}{\small $\sqrt{s}$ [GeV]}
\psfrag{delta}{\small $\overline{\hat{k}_i C_{ij} \hat{n}_j}$ [\%]}
\psfrag{hel+1}{\small $\lambda_-= +$1}
\psfrag{hel0}{\small $\lambda_-=$ 0}
\psfrag{hel-1}{\small $\lambda_-=-$1}                                              
\includegraphics[angle=-90,width=8cm,clip]{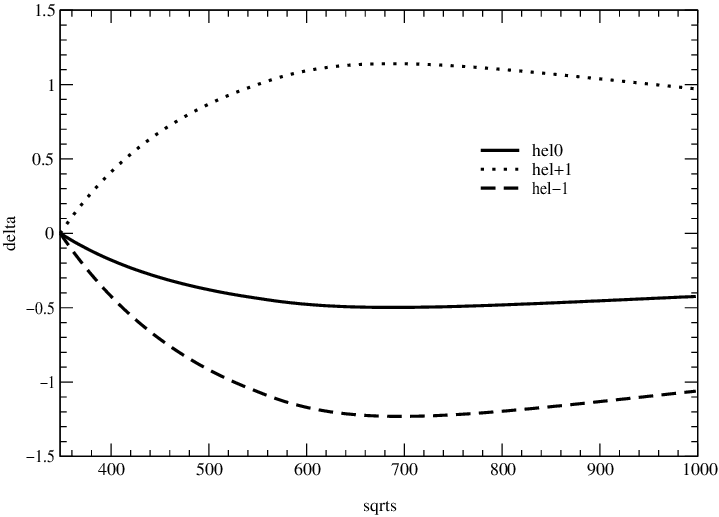}
%
%5a6
\psfrag{delta}{\small $\overline{\hat{p}_i C_{ij} \hat{n}_j}$ [\%]}
\includegraphics[angle=-90,width=8cm,clip]{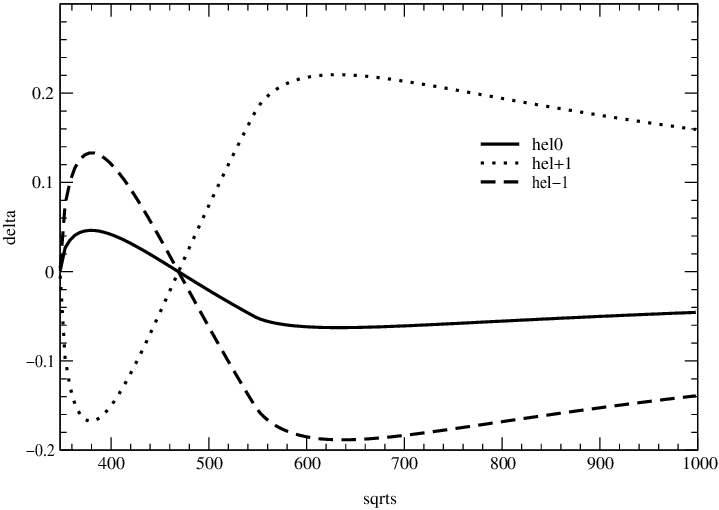}
\caption{Same as Fig.~3, but for the quantities 
$\overline{\hat{k}_i C_{ij} \hat{n}_j}$ (top)  and
$\overline{\hat{p}_i C_{ij} \hat{n}_j}$ (bottom).}
\end{center}
\end{figure}
\section{Conclusions}
In this paper we have derived analytic expressions for the SUSY QCD
corrections to the polarization and spin correlations of $t\bar{t}$ pairs
produced in $e^+e^-$ annihilation with longitudinally polarized beams.
The results depend in particular on the gluino mass and the masses 
of the scalar partners of the top quark. The latter masses depend on the
mixing in the stop sector. For maximal mixing, the SUSY QCD corrections
to the cross section are negative and  reach values of about $-1.3$\% 
($-5$\%) for
a gluino mass of 250 GeV (150 GeV) and a light stop mass 
of 74 GeV. For the same choice of parameters, the $t\bar{t}$ 
spin observables typically receive SUSY QCD corrections well below 1\%.
\section*{Acknowledgments}
We would like to thank A. Djouadi and C. Schappacher for helpful
discussions and D. St\"ockinger for comments on the manuscript.
\section*{Appendix}
\renewcommand{\theequation}{A.\arabic{equation}}
\setcounter{equation}{0}
Here we list explicit results for the formfactors defined
in Eqs. (\ref{hc}), (\ref{ff}). 
Apart from the Standard Model parameters $m_t$, $\alpha_s$, and electroweak
couplings defined in section 3, the formfactors depend on  
the gluino mass $m_{\tilde{g}}$, the 
masses of the two physical top squarks $m_{\tilde{t}_{1,2}}$, 
and the mixing angle $\theta_{\tilde{t}}$ that 
determines how 
the top squark mass 
eigenstates are related to the weak eigenstates, cf. Eq.~(\ref{mix}).

We find (with $C_F=(N_C^2-1)/2/N_C=4/3$):
\bea
V^1_{\gamma}=\frac{\alpha_s}{2\pi}C_FQ_t\left[C_{24}^{11}+
C_{24}^{22}\right]+Q_t\frac{\delta Z_R+\delta Z_L}{2},
\eea
\bea
V^1_{Z}&=&\frac{\alpha_s}{\pi}C_F\left[
\left(g_A^t\cos^2\theta_{\tilde{t}}-Q_t \sin^2\vartheta_W\right)
C_{24}^{11}+
\left(g_A^t\sin^2\theta_{\tilde{t}}-Q_t \sin^2\vartheta_W\right)
C_{24}^{22}\right]
\nonumber \\ &+&
g_V^t\frac{\delta Z_R+\delta Z_L}{2}
-g_A^t\frac{\delta Z_R-\delta Z_L}{2},
\eea

\bea
A^1_{\gamma}=\frac{\alpha_s}{2\pi}C_FQ_t\left[C_{24}^{11}-
C_{24}^{22}\right]\cos 2\theta_{\tilde{t}}
-Q_t\frac{\delta Z_R-\delta Z_L}{2},
\eea
\bea
A^1_{Z}&=&\frac{\alpha_s}{2\pi}C_F\Bigg\{
2\left[
\left(g_A^t\cos^2\theta_{\tilde{t}}-Q_t\sin^2\vartheta_W\right)
C_{24}^{11}-
\left(g_A^t\sin^2\theta_{\tilde{t}}-Q_t\sin^2\vartheta_W\right)
C_{24}^{22}\right]\cos 2\theta_{\tilde{t}}
\nonumber \\ &+&
g_A^t\sin^2 2\theta_{\tilde{t}}\left(C_{24}^{12}+C_{24}^{21}\right)
\Bigg\}
-g_V^t\frac{\delta Z_R-\delta Z_L}{2}
+g_A^t\frac{\delta Z_R+\delta Z_L}{2},
\eea
\bea
S_{\gamma}= \frac{\alpha_s}{2\pi}C_FQ_tm_t\left[s_{\gamma}^{11}
+s_{\gamma}^{22}\right],
\eea
where
\bea
s_{\gamma}^{11(22)}=m_t\left(C_{11}^{11(22)}+C_{21}^{11(22)}\right)
\pm m_{\tilde{g}}\sin 2\theta_{\tilde{t}}
\left(C_{0}^{11(22)}+C_{11}^{11(22)}\right),
\eea
\bea
S_Z&=&\frac{\alpha_s}{2\pi}C_Fm_t\Bigg\{
2\left(g_A^t\cos^2\theta_{\tilde{t}}-Q_t \sin^2\vartheta_W\right)
s_{\gamma}^{11}+
2\left(g_A^t\sin^2\theta_{\tilde{t}}-Q_t \sin^2\vartheta_W\right)
s_{\gamma}^{22}
\nonumber \\ &-&
g_A^t\sin 2\theta_{\tilde{t}}
\cos 2\theta_{\tilde{t}}m_{\tilde{g}}
\left(C_{0}^{12}+C_{11}^{12}+C_{0}^{21}+C_{11}^{21}\right)
\Bigg\}.
\eea
In the above expressions, the one-loop integrals $C_0^{ij},\ldots
C_{24}^{ij}$ are defined by the decomposition of Passarino and Veltman 
\cite{PaVe79},
\bea
C^{ij}_{0;\mu;\mu\nu}=
\frac{(2\pi\mu)^{4-d}}{i\pi^2}\int d^d l\frac{1;l_{\mu};l_{\mu}l_{\nu}}
{[l^2-m^2_{\tilde{g}}+i\epsilon][(l-k_t)^2-m^2_{\tilde{t}_i}+i\epsilon]
[(l+k_{\bar{t}})^2-m^2_{\tilde{t}_j}+i\epsilon]}
\eea
with ($k_V=k_t+k_{\bar{t}}$):
\bea
C^{ij}_{\mu}&=&-k_{t\mu}C_{11}^{ij}+k_{V\mu}C_{12}^{ij},\nonumber \\
C^{ij}_{\mu\nu}&=& k_{t\mu}k_{t\nu}C_{21}^{ij}+
k_{V\mu}k_{V\nu}C_{22}^{ij}-
\left(k_{t\mu}k_{V\nu}+ k_{t\nu}k_{V\mu}\right)C_{23}^{ij}
+g_{\mu\nu}C_{24}^{ij}.
\eea
The quantities $\delta Z_{R,L}$ denote the 
one-loop renormalization constants for 
the chiral components of the top quark field in the on-shell renormalization
scheme. They are given explicitly by
\bea
\delta Z_{L(R)}&=& \frac{\alpha_sC_F}{4\pi}\Big\{
2m_t^2\left[(B_1^{1})'+(B_1^{2})'\right]
+2m_{\tilde{g}}m_t\sin 2\theta_{\tilde{t}}\left[(B_0^{1})'-(B_0^{2})'\right]
\nonumber \\ &+& B_1^1+B_1^2\pm\cos  2\theta_{\tilde{t}}
\left[B_1^1-B_1^2\right]
\Big\},
\eea
where 
\bea
B_{0;\mu}^i=\frac{(2\pi\mu)^{4-d}}{i\pi^2}
\int d^d l\frac{1;l_{\mu}}
{[l^2-m^2_{\tilde{g}}+i\epsilon][(l-k_t)^2-m^2_{\tilde{t}_i}+i\epsilon]}=
B_{0}^i;(-k_{t\mu})B_{1}^i,
\eea
and 
\bea
(B_{0(1)}^{i})'=\frac{dB_{0(1)}^i}{dk_t^2}|_{k_t^2=m_t^2}.
\eea


\begin{thebibliography}{99}
\bibitem{tesla} J.A. Aguilar-Saavedra et al., 
TESLA technical design report part III: Physics at an $e^+e^-$ linear
collider [hep-ph/0106315].
\bibitem{susy} H.P. Nilles, Phys. Rep. {\bf 110} (1984) 1;
H.E. Haber, G. Kane, Phys. Rep. {\bf 117} (1985) 75.
\bibitem{DjDrKo93} A. Djouadi, M. Drees, H. K\"onig, Phys. Rev. D {\bf 48}
(1993) 3081 [hep-ph/9305310].
\bibitem{HoSc98} W. Hollik, C. Schappacher,  Nucl. Phys. B {\bf 545} (1999) 98
[hep-ph/9807427].
\bibitem{KuReZe86} J.H. K\"uhn, Nucl. Phys. B {\bf 237} (1984) 77; 
 J.H. K\"uhn, A. Reiter, P.M. Zerwas, Nucl. Phys. B {\bf 272} (1986) 560.  
\bibitem{BrFlUw99} 
A. Brandenburg, M. Flesch, and  P. Uwer,
Phys. Rev. D {\bf 59} (1999) 014001 [hep-ph/9806306]; 
Chechoslovak Journal of Physics, v. 50 (2000) Suppl. S1, 51-58
[hep-ph/9911249].
\bibitem{schmidt} C. Schmidt, Phys. Rev. D {\bf 54} (1996) 3250 
[hep-ph/9504434].
 \bibitem{epeman1} 
J.G. K\"orner, A. Pilaftsis, and M.M. Tung, Z. Phys. C {\bf 63} 
(1994) 575 [hep-ph/9311332]; 
M.M. Tung, Phys. Rev. D {\bf 52} (1995) 1353 [hep-ph/9403322];
S. Groote and J.G. K\"orner,  
Z. Phys. C {\bf 72} (1996) 255 [hep-ph/9508399]; 
V. Ravindran, W.L. van Neerven, Nucl. Phys. B589 (2000) 507 
[hep-ph/0006125]. 
\bibitem{epeman2} M.M. Tung,  J. Bernabeu,  and J. Penarrocha,
Phys. Lett. B {\bf 418} (1998) 181 [hep-ph/9706444]; 
S. Groote, J.G. K\"orner, and J.A. Leyva,
Phys. Lett. B {\bf 418} (1998) 192 [hep-ph/9708367].
\bibitem{Stu91}
R. G. Stuart, Phys. Lett. B {\bf 262}, 113 (1991);
A. Aeppli, G. J. van Oldenborgh and D. Wyler, Nucl. Phys. B {\bf 428}, 126
(1994).
\bibitem{BeBeCh99} W.~Beenakker, F.A. Berends and A.P. Chapovsky,
Phys.\ Lett.\ B\ {\bf 454}, 129 (1999) [hep-ph/9902304].
\bibitem{BeBrSiUw01} W.~Bernreuther, A. Brandenburg, Z.G. Si, P. Uwer,
Phys. Rev. Lett. {\bf 87} (2001) 242002 [hep-ph/0107086].
\bibitem{CzJeKu91}
M. Jezabek and J. H. K\"uhn, Nucl. Phys. B {\bf 320}, 20 (1989);
A.~Czarnecki, M.~Jezabek and J.~H.~K\"uhn,
Nucl.\ Phys.\ B {\bf 351}, 70 (1991).
\bibitem{BrMa02} A. Brandenburg, M. Maniatis, Phys. Lett. B {\bf 545} (2002)
139 [hep-ph/0207154].
\bibitem{Sc02} C. Schappacher, Ph.D. thesis, Univ. Karlsruhe 2002 
(unpublished); 
W. Hollik, J.I. Illana, S. Rigolin, C. Schappacher, 
D. Stöckinger, Nucl.Phys. B {\bf 551} (1999) 3; Erratum-ibid. B {\bf 557} 
(1999) 407 [hep-ph/9812298].
\bibitem{PaSh96} S. Parke, Y. Shadmi, Phys. Lett. B {\bf 387} (1996) 199  
[hep-ph/9606419].
\bibitem{aleph} ALEPH collaboration, CERN-EP/2002-026, hep-ex/0204036.
\bibitem{scha_pr} C. Schappacher, private communication.
\bibitem{PaVe79} G. Passarino, M. Veltman, Nucl. Phys. B {\bf 160} (1979) 151.
\end{thebibliography}
\end{document}